\documentstyle[psfig,amssymb]{mn}
 
\def\lunits{$\rm erg~s^{-1}$}
\def\funits{$\rm erg~cm^{-2}~s^{-1}$}
\def\cunits{$\rm cm^{-2}$}
\def\xmm{{\it XMM-Newton~}}
\def\chandra{{\it Chandra~}}

\title[The galaxy X-ray luminosity function]{The XMM-Newton Needles in
the Haystack Survey: the local X-ray luminosity function of `normal'
galaxies}   
 
\author[Georgantopoulos et al.] {I. Georgantopoulos$^1$,
  A. Georgakakis$^1$, E. Koulouridis$^{1,2}$ 
  \\ \\
 $^1$Institute of Astronomy \& Astrophysics, National Observatory of
  Athens, I. Metaxa \& V. Pavlou, Athens, 15236, Greece \\ 
 $^2$Astronomical Laboratory, Department of Physics, University of
  Patras, 26500 Rio-Patras, Greece\\
}
\begin{document}
\maketitle  

\begin{abstract}
 In this paper we estimate the local ($z<0.22$) X-ray luminosity 
 function of `normal' galaxies derived from the \xmm Needles in the
 Haystack Survey. This is an on-going project that aims to identify
 X-ray selected `normal' galaxies (i.e. non-AGN dominated) in the
 local Universe. We are using a total of 70 \xmm fields covering an area
 of 11 $\rm deg^2$ which overlap with the Sloan Digital Sky Survey
 Data Release-2. `Normal' galaxies are selected on the basis of their
 resolved optical light profile, their low X-ray--to--optical flux
 ratio ($\log ( f_x / f_o ) < -2)$ and soft X-ray colours. We find a
 total of  28 candidate `normal' galaxies to the 0.5-8\,keV band flux
 limit of $\approx2\times10^{-15}$ \funits. Optical spectra are
 available for most sources in our sample (82 per cent). These provide 
 additional evidence that our sources are bona-fide 'normal' galaxies
 with X-ray emission coming from diffuse hot gas emission and/or X-ray
 binaries  rather than a supermassive black hole.  Sixteen of our galaxies
 have  narrow emission lines or a late-type Spectral Energy
 Distribution (SED)  while the remaining 12 present only absorption
 lines or an early-type SED. Combining our \xmm sample with 18 local
 ($z<0.22$) galaxies from the {\it Chandra} Deep Field North and South
 surveys, we construct  the local X-ray luminosity function of
 `normal' galaxies. This can be represented with a Schechter form  
 with a break at $\rm L_\star\approx 3^{+1.4}_{-1.0}\times10^{41} 
 \rm \, erg \, s^{-1}$ and  
 a slope of $\alpha\approx 1.78\pm 0.12$. Using this luminosity function 
 and assuming pure luminosity evolution  of the form $\propto
 (1+z)^{3.3}$ we estimate a contribution to the X-ray background from
 `normal' galaxies of $\sim$ 10-20 per cent (0.5-8\,keV). 
 Finally,  we derive, for the first time, the luminosity 
 functions  for early and late type systems separately.  
\end{abstract}


\section{Introduction}
 In the last decade, `normal' galaxies (i.e. non-AGN dominated) have
 been studied in detail at X-ray wavelengths by various missions
 through observations of optically selected systems
 (e.g. Fabbiano, Kim \& Trinchieri 1992; Read, Strickland \& Ponman
 1997; Kilgard et al. 2002). The X-ray
 emission in these systems appears to come from diffuse hot gas and/or
 X-ray binaries.  In the most massive early-type systems the X-ray
 emission is dominated by  the hot interstellar medium having
 temperatures of about 1\,keV. Low mass X-ray binaries associated
 with the older stellar population are responsible for a smaller
 fraction of the observed X-ray luminosity. In late-type systems,  the
 X-ray emission originates in both hot gas with  temperature of about
 $\rm kT\sim 1$\,keV (heated by supernova remnants),  as well as a
 mixture of low and high mass X-ray binaries (for a review 
 see Fabbiano 1989). The diffuse
 hot gas contributes significantly in the  soft X-ray band ($<$2 keV)
 while the X-ray binary systems are responsible for the bulk of the
 emission at harder energies (e.g. Stevens, Read \& Bravo-Guerrero
 2003). In general, the integrated X-ray emission of `normal' galaxies
 is believed to be a good indicator of the star-formation activity in
 these systems (e.g. Gilfanov,  Grimm \& Sunyaev 2004). 

 The X-ray luminosity of `normal' galaxies is usually weak, 
 $\la 10^{42}$ \lunits a few orders of
 magnitude below that of powerful AGNs, resulting in faint observed X-ray
 fluxes. As a consequence, until recently, only the very local
 systems ($\rm <100 \, Mpc$) had been accessible to X-ray missions
 leaving the issue of galaxy evolution at X-ray wavelengths open.     
 This has completely changed with the new generation X-ray missions,
 the {\it Chandra} and the {\it XMM-Newton}. The \chandra Deep Fields
 (CDF; Alexander et al. 2003; Giacconi et al. 2002) reaching fluxes
 $f (\rm 0.5-2\, keV)<10^{-16} \, erg \, s^{-1} \, cm^{-2}$ have
 indeed, provided the  
 first ever X-ray selected `normal' galaxy sample. In a pioneering
 work, Horneschemeier et al. (2003) used the 2\,Ms CDF-North to
 provide a sample of 43 `normal' galaxy candidates with available
 optical  spectroscopy.  Norman et al. (2004) extended this study and 
 identified over 100 `normal' galaxy candidates  in the combined
 CDF-North and South albeit with optical spectroscopy limited to 
 a fraction of them. Both these studies find distant galaxies
 ($z<1$) at a median  redshift of $z \approx 0.3$. 
 
 Despite the great progress in the field achieved by {\it Chandra}, there 
 is a pressing need for a local X-ray selected `normal' galaxy
 sample to complement the deeper CDF studies and to provide the X-ray 
 luminosity function of these systems in the nearby
 Universe. Motivated by this we initiated a project using  \xmm
 aiming to identify X-ray selected `normal' galaxies at bright
 fluxes. The large field-of-view combined with the high effective area  
 of \xmm make this mission ideal for this study.  Our fields are
 selected to overlap with the Sloan Digital Sky Survey  (SDSS),  Data
 Release-2 to exploit the good quality and homogeneous five-band
 optical photometry and optical spectroscopy. First
 results from this on-going survey have been reported in Georgakakis
 et al. (2004b). Similar studies have been 
 recently performed with the HRI detector onboard {\it ROSAT}
 (Tajer et al. 2005). 
 Our main goal in  this paper is to expand the Georgakakis
 et al. (2004b) sample to determine the X-ray galaxy luminosity   
 function in the local Universe, $z\la0.2$.
 Throughout this paper we
 adopt $\rm Ho = 70 \, km \, s^{-1} \, Mpc^{-1}$, $\rm \Omega_{M} =
 0.3$ and $\rm  \Omega_{\Lambda} = 0.7$.      

\section{The Data Acquisition}

\subsection{The XMM-Newton Observations} 

In this paper we use {\it XMM-Newton} archival observations, with a
proprietary period that expired before June 2004, that overlap
with second data release of the SDSS (DR2; Stoughton et al. 2002). Only
observations that use the EPIC (European Photon Imaging Camera; 
Str\"uder et al. 2001; Turner et al. 2001) cameras as the prime
instrument operated in full frame mode were employed. We use
only fields at high Galactic latitude $|b|>20^\circ$ in order 
 to minimize the absorption as well as the stellar contamination. 
 We also reject fields  which contain bright clusters as their target. 
 Finally, fields that are heavily
contaminated by high particle background periods are  excluded from
the analysis. For fields observed more than once with the {\it
XMM-Newton} we use the deeper of the multiple observations. A total of
42 new fields are used in addition to the 28 {\it XMM-Newton}
observations used in  Georgakakis et al. (2004b). Details of all   
70 fields are given in  Table \ref{log}.

We are using the  Pipeline Processing Subsystem (PPS) event files.  
 The event files were screened for
high particle  background periods by rejecting times with 0.2-10\,keV
count rates higher than 25 and 15\,cts/s for the PN and the two MOS
cameras respectively. The resulting PN and MOS exposures 
 are shown in Table \ref{log}. The
differences between the PN and MOS exposure times are due to varying
start and end times of individual observations. Only events 
corresponding to patterns  0--4 for the PN and 0--12 for two MOS
cameras have been kept.
 To increase the signal--to--noise ratio and to
reach fainter fluxes the PN and the MOS event files,
 where available,  have been combined
into a single event list using the {\sc merge} task of SAS.
Images in celestial coordinates with pixel size of 4.35\,arcsec have
been extracted in the spectral bands 0.5-2\,keV (soft) and 2-8\,keV
(hard) for the merged event file. Exposure maps accounting for
vignetting, CCD gaps and bad pixels  have been constructed for each
spectral band. We apply no astrometric corrections in our data.
 However, we estimate the astrometric accuracy of the 
 {\it XMM-Newton} positions to be better than 3 arcsec (see section 3).  
Source detection is performed in the 0.5-8\,keV merged 
PN+MOS images using the {\sc ewavelet} task of {\sc sas} with a
detection threshold of $5\sigma$.  
Count rates in the merged (PN+MOS) images as well as the
individual PN and MOS images are estimated within an 18\,arcsec
aperture. For the  background estimation we use the background maps
generated as a by-product of the {\sc ewavelet} task of  {\sc SAS}. The
merged image count rates are used for flux estimation, while the
individual PN or MOS count rates are used for hardness ratios. This is
because the interpretation of hardness ratios is simplified if the
extracted count rates  are from one detector only. A small fraction of
sources lie close to masked regions (CCD gaps or hot pixels) on either
the MOS or the PN detectors. This may introduce errors in the
estimated source counts. To avoid this bias, the source count rates
(and hence the hardness ratios and the flux) are estimated  using the
detector (MOS or PN) with no masked pixels in the vicinity of the
source.  We define the hardness ratio as HR=H-S/H+S, where H and S are
the source count rates in the 2-8 keV and 0.5-2 keV band
respectively. Hence, a more negative hardness ratio value, suggests a
softer (steeper) X-ray spectrum. To convert counts to flux the Energy
Conversion Factors (ECF) of individual detectors are calculated
assuming a power law spectrum with $\Gamma=1.8$ and Galactic
absorption  appropriate for each field (Dickey \& Lockman 1990). We
do not apply any correction for Galactic absorption. However, for the
median Galactic column  density  of our `normal' galaxy sample
($N_H\approx  2.8\times10^{20}$ \cunits) such a correction is small,
only about 4 per cent of the 0.5-8 keV flux.

\begin{figure}
\centerline{\psfig{figure=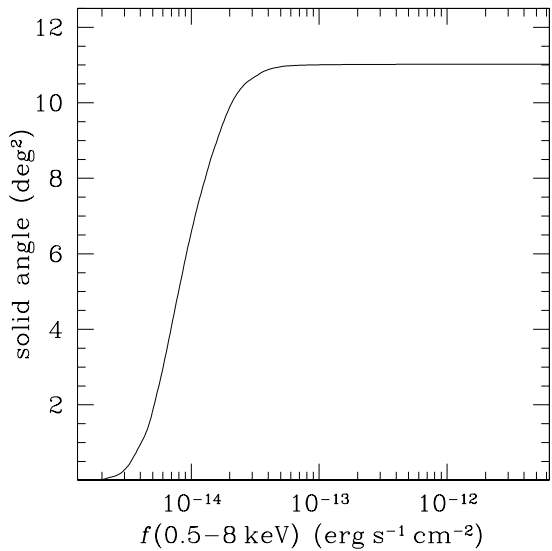,width=3in,angle=0}}
\caption
 {Solid angle as a function of limiting flux ($5\sigma$) in the
total 0.5-8\,keV band for our survey. 
 }\label{area}
\end{figure}

 Next, we derive the area curve i.e. the solid angle 
 as a function of the 5$\sigma$ limiting flux  for our observations.
 For each pixel of the background map,
 generated as a by product of the {\sc ewavelet} task, we estimate the
 $5\sigma$  background fluctuations. We then scale to the area of a
 circular aperture with a size of 4 pixel radius. The 4 pixel scale
 encircles 70 per cent of the light (at an energy of 1.5keV) and roughly 
 corresponds to the scale of the wavelet filter used for detection. 
 These values are then divided with the corresponding exposure time
 and are converted to flux. The area curve is derived using the merged
 PN and MOS background and exposure maps where available or the single
 PN and MOS maps. We have checked that the area curve derived above
 gives reasonable results by estimating the 0.5-8\,keV $\log N -\log
 S$ for all the X-ray sources in our fields and comparing with the
 number counts derived from other surveys (e.g. Manners et
 al. 2003). Figure \ref{area} shows the solid angle covered  by our
 survey as a function of the  0.5-8\,keV limiting flux.

\subsection{The SDSS data} 

The SDSS is an ongoing imaging and spectroscopic survey which has
covered so far (DR-2) $\approx \rm 3324 \, deg^2$ of the
sky. Photometry is  performed in 5 bands  ($ugriz$;  Fukugita et
al. 1996; Stoughton et al. 2002) to the limiting magnitude $g \approx
23$\,mag, providing a uniform and homogeneous multi-color photometric
catalogue. The SDSS spectroscopic observations obtain spectra of
galaxies brighter than $r=17.7$\,mag as well as of luminous red
galaxies with $r<19.2$\,mag (York et al. 2000; Stoughton et al. 2002).    

We identify the optical counterparts of the X-ray sources following 
the method of Downes et al. (1986)  to calculate the probability, $P$,
that a given candidate is the true identification. We apply an upper
limit in the search radius, $r<7\rm\,arcsec$ and a cutoff in the
probability, $P<0.05$, to limit the optical  identifications to those
candidates that are least likely to be chance coincidences. 

\begin{table*}
\footnotesize
\begin{center}
\begin{tabular}{cc cc cc cc}
\hline
 Obs ID & RA  & Dec  & FILTER & $\rm N_H$  & PN exp. time & MOS1 exp. time &
 Field name \\
& (J2000) & (J2000) & & ($\rm 10^{20}\,cm^{-2}$)  & (ksec) &  (ksec) &  \\
\hline
0065140101 & 00 42 31 & -09 41 29 & MEDIUM & 3.60 & 9.0 & 12.4 & ABELL85  \\
0090070201 & 00 43 20 & --00 51 15 & MEDIUM & 2.33 & 15.7 &  --  & UM\,269 \\
0093641001 & 01 43 02 & +13 38 30 & MEDIUM & 4.87 & 6.3 & 11.0 & NGC 660 \\
0084230401 & 01 52 42 & +01 00 43 & MEDIUM & 2.80 & 5.8  & 17.2 & ABELL\,267 \\
0101640201 & 01 59 50 & --00 23 41 & MEDIUM & 2.65 & 3.8  &  --     & MRK\,1014 \\
0093630101 & 02 41 05 & --08 15 21 & MEDIUM & 3.07 & 12.3 & 15.6 & NGC\,1052 \\
0056020301 & 02 56 33 & --00 06 12 & THIN   & 6.50 &  --     & 11.6 & RX\,J0256.5+0006 \\
0041170101 & 03 02 39 & --00 07 40 & THIN   & 7.16 & 38.1 & 46.9 & CFRS\,3H \\
0142610101 & 03 06 41 & +00 01 12 & THIN & 6.96 & 33.5 & 45.8 & S2F1A  \\
0103861001 & 03 25 25 & -06 08 30 & MEDIUM & 4.39 & 6.8 & - & MRK609  \\
0134540601 & 03 36 47 & +00 35 15 & MEDIUM & 8.17 & 30.3 & 35.1 & HR1099 \\
0036340101 & 03 38 29 & +00 21 56 & THIN   & 8.15 & 8.9  & 6.7  & SDSS\,033829.31+00215 \\
0094790201 & 03 57 22 & +01 10 56 & THIN   &13.20 & 19.1 & 21.4 & HAWAII\,167 \\
0152530101 & 08 10 57 & +28 08 33 & THIN & 3.73 & 16.7 & 22.1 & YZ CNC \\
0092800201 & 08 31 41 & +52 45 18 & MEDIUM & 3.83 & 66.8 & 73.3 & APM\,08279+5255 \\
0111971701 & 08 38 22 & +48 38 01 & MEDIUM & 3.41 & 7.2 & - & EI UMA \\
0103660201 & 08 47 42 & +34 45 05 & MEDIUM & 3.28 & 13.0 & - & PG 0844+349 \\
0085150301 & 08 49 18 & +44 49 24 & MEDIUM & 2.63 & 29.3 & 35.1 & LYNX 3A-SE  \\
0083240201 & 09 11 27 & +05 50 52 & THIN & 3.67 & - & 17.7 & RX J0911.4 \\
0084230601 & 09 17 53 & +51 43 38 & MEDIUM & 1.44 & 15.9 & 13.6 & ABELL\,773 \\
0112520201 & 09 34 02 & +55 14 20 & THIN   & 1.98 & 23.5 & 28.5 & IZW\,18 \\
0085640201 & 09 35 51 & +61 21 11 & THIN   & 2.70 & 20.4 & 33.9 & UGC\,5051 \\
0106460101 & 09 43 00 & +46 59 30 & THIN & 1.24 & - & 46.5 & CL0939+472  \\
0070940401 & 09 53 41 & +01 34 46 & THIN & 3.46 & - & 12.9 & NGC3044  \\
0070340201 & 10 08 48 & +53 42 03 & THIN & 7.57 & 18.0 & 20.8 & WJ1008.7  \\
0108670101 & 10 23 40 & +04 11 24 & THIN & 2.94 & 45.6 & 52.6 & ZW 3146 \\
0147511701 & 10 52 41 & +57 28 29 & MEDIUM & 5.58 & 84.1 & 95.5 & LOCKMAN HOLE \\
0083000301 & 11 23 09 & +05 30 19 & MEDIUM & 4.39 & 23.3 & 28.2 & 3C 257 \\
0112810101 & 11 28 30 & +58 33 43 & THIN & 9.92 & 13.8 & 20.0 & NGC 3690 \\
0111970701 & 11 38 27 & +03 22 07 & MEDIUM & 2.36 & 9.0 & - & T LEO \\
0094800201 & 11 40 23 & +66 08 41 & THIN & 1.18 & 19.0 & 24.7 & MS1137.5  \\
0044740201 & 11 50 42 & +01 45 53 & THICK & 2.22 & 41.3 & 47.9 & BETA VIR \\
0049340301 & 11 51 07 & +55 04 45 & MEDIUM & 1.14 & 20.3 & 25.0 & NGC 3921 \\
0090020101 & 11 57 56 & +55 27 12 & THIN & 1.22 & 8.0 & 11.1 & NGC 3998 \\
0056020701 & 12 00 48 & -03 27 51 & THIN & 2.35 & 22.1 & 29.0 & RXJ1200.8  \\
0081340801 & 12 13 46 & +02 48 41 & THIN & 1.78 & 17.8 & 22.3 & IRAS12112  \\
0056340101 & 12 19 23 & +05 49 31 & MEDIUM & 1.56 & 22.0 & 27.5 & NGC4261  \\
0110990201 & 12 27 19 & +01 29 24 & THIN & 1.85 & 7.9 & 9.7 & HI1225+01  \\
0124900101 & 12 31 32 & +64 14 21 & THIN   & 1.98 & 26.1 & 30.1 & MS\,1229.2+6430 \\
0111550401 & 12 36 57 & +62 13 30 & THIN & 1.51 & 75.4 & 87.1 & HUBBLE DEEP \\
0110980201 & 12 45 09 & --00 27 38 & MEDIUM & 1.73 & 46.3 & 55.5 & NGC\,4666 \\
0136000101 & 13 04 12 & +67 30 25 & THIN   & 1.80 & 14.6 & 17.1 & ABELL\,1674 \\
0056021001 & 13 08 33 & +53 42 19 & THIN & 1.53 & 22.2 & 28.0 & RX J1308.5 \\
0111281601 & 13 41 24 & --00 24 00 & THIN &1.4& 3.5 & 7.1   &F864-7\\
0111281001 & 13 41 24 & +00 24 00 & THIN & 1.0 & 5.8  & 10.0 & F864-1\\
0111281401 & 13 43 00 & +00 00 00 &THIN & 2.8 & 1.7 & 4.5  & F864-5 \\
0111282401 & 13 43 00 & +00 24 00 & THIN& 2.0 & 3.0  & 6.6 &F864-2\\
0111281701 & 13 43 24 & --00 24 00 & THIN&2.2 & 2.1 & 7.3   &F864-8\\
0111281801 & 13 44 36 & --00 24 00 & THIN & 3.6& -- & 7.7   &F864-9 \\   
0111281501 & 13 44 36 & +00 00 00 & THIN & 1.6 & 2.8 & 6.5    &F864-6\\
0111282601 & 13 44 36 & +00 24 00 & THIN& 2.0 & 2.2  & 7.7 &F864-3\\
0112250201 & 13 47 41 & 58 12 42 & MEDIUM & 1.28 & 24.7 & 31.3 & QSO 1345+584 \\
0071340501 & 13 49 15 &  +60 11 26 & THIN   & 1.80 & 14.1 & 18.1 & NGC\,5322 \\
0112250101 & 13 54 17 & -02 21 46 & THIN & 3.32 & 20.4 & 24.2 & RXJ1354.3  \\
0110930401 & 14 35 30 & 48 44 30 & MEDIUM & 2.08 & - & 7.0 & NGC5689  \\
0021540101 & 15 06 29 & 01 36 20 & THIN & 4.24 & 25.7 & - & NGC 5846 \\
0111260201 & 15 10 03 & 57 02 44 & THIN & 1.49 & 8.4 & 11.4 & GB1508+5714  \\

\hline
\end{tabular}
\end{center}
\caption{The {\it XMM-Newton} pointings}
\label{log}
\normalsize
\end{table*}

\begin{table*}
\contcaption{}
\footnotesize
\begin{center}
\begin{tabular}{cc cc cc cc}
\hline
 Obs ID & RA  & Dec  & FILTER & $\rm N_H$  & PN exp. time & MOS1 exp. time &
 Field name \\
& (J2000) & (J2000) & & ($\rm 10^{20}\,cm^{-2}$)  & (ksec) &  (ksec) &  \\
\hline
0145190201 & 15 15 54 & 56 19 44 & THIN & 1.44 & 19.4 & 28.6 & NGC 5907 \\
0103860601 & 15 16 40 & 00 14 54 & THICK & 4.67 & 8.9 & 13.2 & CGCG21-63  \\
0011830201 & 15 25 54 & 51 36 49 & THIN & 1.56 & 24.7 & 29.7 & CSO 755 \\
0150610301 & 15 36 38 & 54 33 33 & THIN & 1.32 & 16.0 & 24.3 & PG 1535+547 \\
0060370901 & 15 43 59 &  +53 59 04 & THIN   & 1.27 & 14.2 & 19.2 & SBS\,1542+541 \\
0025740401 & 16 04 19 & 43 04 33 & THIN & 1.25 & 12.5 & 15.7 & CL1604+4304  \\
0033540901 & 16 32 01 & 37 37 50 & THIN & 1.17 & 11.0 & 14.3 & PG 1630+377 \\
0107860301 & 17 01 23 &  +64 14 08 & MEDIUM & 2.65 & 2.3  & 3.9  & RXJ\,1701.3 \\
0111180201 & 20 40 10 & -00 52 16 & MEDIUM & 6.70 & 8.7 & - & AE AQR \\
0093030201 & 21 29 38 & 00 05 38 & MEDIUM & 4.29 & 29.0 & 40.0 & RXJ2129.6  \\
0042341301 & 23 37 40 & --00 16 33 & THIN   & 3.82 & 8.2  & 13.3 & RXCJ\,2337.6+0016 \\
0147580401 & 23 47 25 & 00 53 58 & THIN & 3.77 & 12.2 & 15.0 & 1AXGJ234725  \\
0108460301 & 23 54 09 & --10 24 00 & MEDIUM & 2.91 & 13.6 & 19.1 & ABELL\,2670 \\
\hline
\end{tabular}
\end{center}
\normalsize
\end{table*}

\section{Galaxy selection}
`Normal' galaxy candidates are selected to have  (i) extended optical
light profile, i.e. resolved (see Stoughton et al. 2002),
  to avoid contamination of the sample by Galactic stars, 
 (ii) X-ray--to--optical flux ratio $\log (f_x / f_o)
< -2$, two orders of magnitude lower than typical AGNs. The
$\log f_X / f_{o}$ is estimated 
from the relation
\begin{equation}\label{eq2}
\log\frac{f_X}{f_{o}} = \log f_X(0.5-8\,{\rm keV}) +
0.4\,r + 5.39.
\end{equation}
The equation above is derived from the X-ray--to--optical flux
ratio definition of Stocke et al. (1991) that involved 0.3-3.5\,keV
flux and $V$-band magnitude. These quantities are converted to
0.5-8\,keV flux and $r$-band magnitude assuming a mean colour
$V-R=0.7$ and a power-law X-ray spectral energy distribution with
index  $\Gamma=1.8$. The sample of `normal' galaxy candidates is
presented in Table \ref{sample}.  We further exclude from the sample
sources with hard X-ray colours (hardness ratio  $HR>0$) roughly
corresponding to a spectrum with a hydrogen column density higher than
$\rm 10^{22} \, cm^{-2}$  (assuming  a power-law index of
$\Gamma=1.9$). The three hard sources (\#7, 15, 28) are most likely
associated with low luminosity obscured AGN.  These are presented in
Table \ref{hard}. Note however, that for a few sources (\#08, 10, 14,
29)  we do not have enough photon statistics to place a stringent
constraint on their X-ray spectrum.  We also exploit the SDSS optical
spectroscopic information available for our sources to search for AGN
signatures using emission line ratios. Sixteen of our galaxies have
either a narrow emission line optical spectrum or a Spectral Energy
Distribution (SED), as derived from the SDSS colours, consistent with
a late-type spectrum. Twelve galaxies either present only absorption
lines or their SED is consistent  with an early-type spectrum. We
employ the CMU-PITT SDSS Value Added Database (VAC
database\footnote{\sc  http://astrophysics.phys.cmu.edu/dr3/}) which
provides spectral classifications for the SDSS galaxies using
diagnostic emission line ratios ($\rm [N\,II]/H\alpha$, $\rm
[O\,III]/H\beta$, $\rm [S\,II]/H\alpha$, $\rm [O\,I]/H\alpha$; Miller
et al. 2003).  All emission line systems in Table \ref{sample} with
available spectroscopic classifications have emission-line ratios
consistent with star formation activity. However, we note that a
number of sources with both absorption (e.g. $\rm H\beta$) and
emission-line optical spectra have uncertain classification based on
one line ratio only, usually $\rm [N\,II]/H\alpha$.  These are marked
in Table \ref{sample}. 

The archival X-ray data used here include targeted observations of
nearby normal galaxies with low X-ray--to--optical flux ratio. Such
sources have been excluded from Table \ref{sample}. Moreover, a number
of `normal' galaxy  candidates although not the prime target of the
{\it XMM-Newton} pointing lie at the same redshift as the prime target
and are therefore most likely directly associated with it
(e.g. cluster or group members). These sources are marked in Table
\ref{sample}. The final sample of 'normal' galaxy candidates that are
not showing evidence  for  AGN activity comprises 28 sources. Of these
only  five do not have optical spectroscopy
available. Although no astrometric corrections have been applied to
our data the positional accuracy of {\it XMM-Newton} is sufficient for
the purpose of this paper. We quantify the astrometric accuracy of the
{\it XMM-Newton} by constructing the distribution of the X-ray/optical
position angular offset for all the X-ray sources detected in our
survey with an optical counterpart. This is well described by a
Gaussian distribution with $1\sigma$ rms of 1.7\,arcsec for the bright
X-ray sources ($>$75 counts) and 2.8\,arcsec for the fainter ones
($<$75 counts). In Table  \ref{sample}  the offset between the optical
and X-ray coordinates for our 28 galaxies is  $\la 5$\,arcsec, within
the $2\sigma$ positional uncertainty for faint sources derived above.

\begin{table*} 
\scriptsize
\begin{center} 
\begin{tabular}{l ccc cccc  ccc c}
\hline 
ID &
$\alpha_X$ & 
$\delta_X$ & 
$r$    &
$P$    &
$\delta_{XO}$ &
$f_x$ &
HR   &
$\log (f_x/f_o)$ &
$z$ & 
$\log L_X$ &
type$^1$ \\
 
 &
(J2000) & 
(J2000) & 
(mag)    &
(\%)     &
(arcsec) &
($10^{-14}$\,cgs) &
  &
  &
  & 
($\rm erg\,s^{-1}$) &
  \\

\hline
01 & 00 42 44.68 & $-$09 33 16.27 & 15.20 & $0.01$  & 1.1 & $1.41\pm0.52$ & 
$<-0.22$ & $-$2.38 & $0.054^2$ & 40.99 & A$^{}$ \\ 
02$^{}$ & 03 06 56.92 & $-$00 00 24.41 & 17.71 & $0.16$ & 2.5 & $0.32\pm0.14$ 
& $<-0.70$ & $-$2.02 & $0.109$ & 40.99 & C$^{}$ \\ 
03$^{}$ & 03 25 31.40 & $-$06 07 44.04 & 14.58 & $0.02$ & 2.6 & $1.02\pm0.47$ 
& $<-0.13$ & $-$2.77 & $0.035^{2}$ & 40.46 & E$^{}$ \\ 
04      & 03 58 05.25 & $+$01 09 50.34 & 14.93 & $0.01$ & 1.4 & $1.52\pm0.34$ 
& $-0.46\pm0.28$ & $-$2.45 & $0.074^{3}$ & 41.30 & A$^{}$ \\ 
05$^{}$ & 08 30 59.81 & $+$52 37 47.36 & 17.69 & $0.80$ & 5.4 & $0.34\pm0.12$ 
& $-0.18\pm0.84$ & $-$2.00 & $0.136$ & 41.24 & E$^{}$ \\ 
06$^{}$ & 08 31 14.62 & $+$52 42 25.32 & 15.24 & $0.01$ & 0.9 & $0.56\pm0.09$ 
& $-0.79\pm0.19$ & $-$2.76 & $0.064$ & 40.74 & C$^{}$ \\ 
08$^{}$ & 08 32 02.52 & $+$52 47 12.90 & 16.01 & $0.01$ & 1.0 & $0.25\pm0.08$ 
& $<+0.60$ & $-$2.81 & $0.105^3$ & 40.84 & E$^{}$ \\ 
09$^{}$ & 08 32 28.21 & $+$52 36 22.74 & 13.82 & $<0.01$ & 0.7 &$6.43\pm0.28$ 
& $<-0.99$ & $-$2.27 & $0.017$ & 40.65 & E$^{}$ \\ 
10$^{}$ & 09 17 58.41 & $+$51 51 08.91 & 16.68 & $0.33$ & 4.6 & $0.61\pm0.26$ 
& $<+0.07$ & $-$2.15 & $0.219^{2,3}$ & 41.92 & A$^{}$ \\ 
11$^{}$ & 09 35 18.92 & $+$61 28 34.43 & 16.50 & $0.27$ & 4.7 & $0.95\pm0.24$ 
& $-0.37\pm0.28$ & $-$2.03 & $0.124$ & 41.58 & E$^*$ \\ 
12$^{}$ & 09 36 19.41 & $+$61 27 20.85 & 16.55 & $0.05$ & 2.3 & $0.95\pm0.21$ 
& $<-0.41$ & $-$2.01 & $0.131$ & 41.65 & A$^{}$ \\ 
13$^{}$ & 10 08 15.98 & $+$53 42 15.19 & 16.71 & $0.08$ & 2.3 & $0.73\pm0.21$ 
& $<-0.07$ & $-$2.06 & $0.069$ & 40.93 & E$^{}$ \\ 
14$^{}$& 10 23 06.49 & $+$04 08 04.34 & 15.62 & $0.18$ & 4.5 & $0.29\pm0.14$ 
& $<+0.50$ & $-$2.89 & $0.048$ & 40.20 & C$^{}$ \\ 
16$^{}$& 11 23 05.25 & $+$05 38 40.42 & 15.02 & $<0.01$ & 0.3 & $1.28\pm0.24$ 
& $<-0.25$ & $-$2.49 & $0.049$ & 40.86 & E$^{}$ \\ 
17$^{}$ & 11 28 45.85 & $+$58 35 36.53 & 14.96 & $0.15$ & 5.6 & $1.35\pm0.32$ 
& $-0.38\pm0.60$ & $-$2.49 & $0.059$ & 41.05 & E$^{*}$ \\ 
18$^{}$ & 11 50 32.52 & $+$55 03 28.76 & 14.39 & $<0.01$ & 1.0 & 
$1.52\pm0.20$ & $-0.62\pm0.19$ & $-$2.67 & $0.019^2$ & 40.14 & C$^*$ \\ 
19$^{}$ & 11 50 51.05 & $+$55 08 37.14 & 13.52 & $<0.01$ & 1.9 & 
$1.03\pm0.26$ & $-0.38\pm0.26$ & $-$3.19 & $0.019^2$ & 39.93 & C$^*$ \\ 
20$^{}$ & 12 19 35.76 & $+$05 50 48.30 & 12.67 & $<0.01$ & 2.6 & 
$1.29\pm0.24$ & $-0.80\pm0.26$ & $-$3.43 & $0.008^2$ & 39.26 & A$^{}$ \\ 
21$^{}$ & 12 31 46.84 & $+$64 14 03.34 & 13.01 & $<0.01$ & 3.0 & 
$1.95\pm0.14$ & $-0.38\pm0.13$ & $-$3.11 & $0.002$ & 38.39 & A$^{}$ \\ 
22$^{}$ & 12 32 53.11 & $+$64 08 56.02 & 15.43 & $0.17$ & 5.1 & $1.03\pm0.26$ 
& $<-0.26$ & $-$2.42 & $0.140^3$ & 41.72 & E$^{}$ \\ 
23$^{}$ & 12 44 52.21 & $-$00 25 50.70 & 15.34 & $0.01$ &  1.3 & 
$0.56\pm0.11$ & $<-0.34$ & $-$2.72 & $0.082$ & 40.97 & C$^{}$ \\ 
24$^{}$ & 12 45 32.14 & $-$00 32 05.01 & 13.12 & $0.01$ & 4.2 & $2.83\pm0.19$ 
& $-0.45\pm0.09$ & $-$2.91 & $0.005$ & 39.29 & E$^{}$ \\ 
25$^{}$ & 13 03 01.51 & $+$67 25 20.73 & 16.16 & $0.16$ & 3.7 & $1.38\pm0.34$ 
& $<-0.28$ & $-$2.00 & $0.109^{2,3}$ & 41.62 & A$^{}$ \\ 
26$^{}$ & 15 07 07.69 & $+$01 32 39.26 & 11.65 & $<0.01$ & 2.5 & 
$3.30\pm0.37$ & $-0.69\pm0.11$ & $-$3.43 & $0.009$ & 39.83 & A$^{}$ \\ 
27$^{}$ & 15 09 46.77 & $+$57 00 00.76 & 11.68 & $<0.01$ & 1.7 & 
$6.75\pm0.52$ & $-0.27\pm0.10$ & $-$3.11 & $0.003$ & 39.18 & A$^{}$ \\ 
29$^{}$ & 21 29 33.97 & $+$00 01 35.97 & 16.12 & $0.01$ & 1.0 & $0.35\pm0.23$ 
& $<+0.05$ & $-$2.61 & $0.052$ & 40.35 & A$^{}$ \\ 
30$^{}$ & 23 53 40.52 & $-$10 24 17.79 & 15.07 & $0.02$ &  2.1 & 
$2.52\pm0.51$ & $<-0.62$ & $-$2.18 & $0.074^2$ & 41.52 & A$^{}$ \\ 
31$^{}$ & 23 54 05.71 & $-$10 18 33.07 & 15.70 & $0.05$ & 2.5 & $1.06\pm0.33$ 
& $<-0.06$ & $-$2.30 & $0.073^2$ & 41.14 & A$^{}$ \\ 

\hline

\multicolumn{12}{l}{$^1$A: absorption lines; E: Narrow emission
lines; C: both narrow emission and absorption lines. The $^*$ denotes 
ambiguous SDSS spectral classification.} \\ 

\multicolumn{12}{l}{$^2$source at the same redshift as the target of
the {\it XMM-Newton} pointing.}\\


\multicolumn{12}{l}{$^3$Photometric redshift from the SDSS.}\\

\multicolumn{12}{l}{The columns are: 1: identification number; 2,3:
right ascension and declination of the X-ray source (J2000); 4: optical 
magnitude;}\\   

\multicolumn{12}{l}{5: probability the optical counterpart is a 
spurious alignment; 6: X-ray/optical position offset;}\\

\multicolumn{12}{l}{7: absorbed X-ray flux in
the 0.5-8\,keV spectral band in units of $10^{-14} \, \rm erg \,
s^{-1} \, cm^{-2}$;}\\

\multicolumn{12}{l}{8: Hardness ratio derived in the 0.5-2 and 2-8 keV bands; 
 9: X-ray--to--optical flux ratio; 
10: spectroscopic or  photometric redshift from the SDSS;}\\ 

\multicolumn{12}{l}{11: Logarithmic 0.5-8\,keV X-ray  luminosity in units of
$10^{41}\, \rm erg \, s^{-1}$; 12: spectral type.}\\  
\end{tabular} 
\end{center} 
\caption{
The candidate `normal' galaxy sample. 
}\label{sample} 
\normalsize  
\end{table*}

\begin{table*} 
\scriptsize
\begin{center} 
\begin{tabular}{l ccc cccc  ccc c}
\hline 
ID &
$\alpha_X$ & 
$\delta_X$ & $r$    & $P$    & $\delta_{XO}$ &
$f_x$ & HR   & $\log (f_x/f_o)$ &
$z$ &  $\log L_X$ & type$^1$ \\
& (J2000) &  (J2000) &  (mag)    &
(\%)     & (arcsec) & ($10^{-14}$\,cgs) & & & & ($\rm erg\,s^{-1}$) & \\
 \hline
07 & 08 31 39.11 & $+$52 42 06.87 & 15.70 & $0.02$ & 1.4 & $2.07\pm0.14$ 
& $+0.50\pm0.07$ & $-$2.01 & $0.059$ & 41.23 & E$^{}$ \\ 
15 & 10 52 47.89 & $+$57 36 20.56 & 17.98 & $0.13$ & 2.2 & $0.20\pm0.09$ 
& $+0.22\pm0.62$ & $-$2.10 & $0.118$ & 40.87 & E$^{}$ \\ 
28  & 15 16 27.00 & $+$00 23 03.30 & 15.78 & $0.01$ & 1.1 & 
$1.63\pm0.56$ & $+0.20\pm0.34$ & $-$2.08 & $0.053^{2}$ & 41.04 & A$^{}$ \\ 
\hline
\multicolumn{12}{l}{$^1$A: absorption lines; E: Narrow emission
lines; C: both narrow emission and absorption lines.} \\ 
\multicolumn{12}{l}{$^2$source at the same redshift as the target of
the {\it XMM-Newton} pointing.}\\
\multicolumn{12}{l}{Columns as in table \ref{sample}}\\  
\end{tabular} 
\end{center} 
\caption{
The excluded hard Low Luminosity AGN. 
}\label{hard} 
\normalsize  
\end{table*}

\section{The Luminosity Function} 

\begin{figure}
\centerline{\psfig{figure=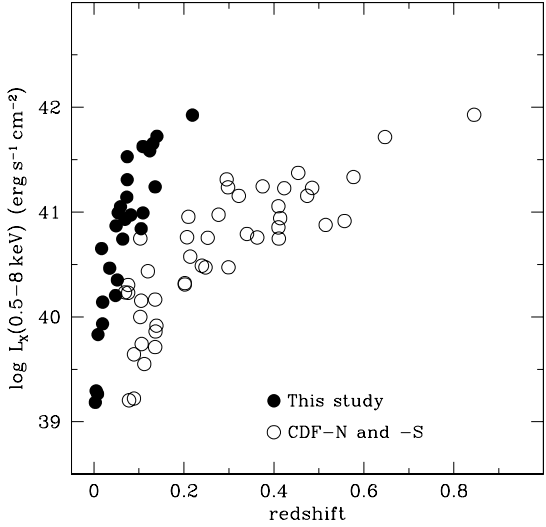,width=3in,angle=0}}
\caption
 {$L_X(\rm 0.5 - 8\,keV)$ against redshift. Filled circles are for our
 sample of `normal' galaxy candidates (including sources associated
 with the prime target of a given {\it XMM-Newton} pointing). Open
 circles represent the CDF samples.  The NHS and CDF 
 surveys cover complementary regions of the $L_X - z$ space.   
}\label{lxz}
\end{figure}

The luminosity-redshift relation for our sample is compared  in
Fig. \ref{lxz} with the `normal' galaxies from the  CDF-North and
South. For the CDF-North  we use the spectroscopic sample of
Hornschemeier  et al. (2003) while for the CDF-South we select $\log
(f_x /f_o)<-2$ sources from the 0.5-2\,keV catalogue of Giaconni et
al. (2002) with spectroscopic or photometric redshifts obtained from
Szokoly et al. (2004) and Zheng et al. (2004). From Fig. \ref{lxz} 
 it can be seen that our sample is complementary to the CDF, 
 covering the  low redshift and
 high luminosity part of the luminosity-redshift plane.   
 
We derive the binned `normal' galaxy X-ray luminosity function using
the method described by Page \& Carrera (2000). This is variant of the
classical non-parametric $1/V_{max}$ method (Schmidt 1968) and has
the advantage that it is least affected by systematic errors for
objects close to the flux limit of the survey. For a given redshift and
X-ray luminosity interval the binned luminosity function is estimated
from the relation: 

\begin{equation}  
\Phi(L)= \frac{N}{\int_{L_{min}}^{L_{max}}  \int_{z_{min}(L)}^{z_{max}(L)}
\Omega(L,z)\,dV/dz\,dz\, dL}, 
\end{equation} 

\noindent 
where $N$ is the number of sources with luminosity in the range
$L_{min}$ and $L_{max}$ and $dV/dz$ is the volume element per redshift
interval $dz$. For a given luminosity $L$,  $z_{min}(L)$ and
$z_{max}(L)$ are the minimum and the maximum redshifts possible for a
source of that luminosity to remain within the flux limits of the
survey and to lie within the redshift bin. $\Omega(L,z)$ is the solid
angle of the X-ray survey available to a source with luminosity $L$ at
a redshift $z$ (corresponding to a given flux in the X-ray area 
curve). The logarithmic bin size of the luminosity function varies so
that each bin comprises approximately equal number of sources $N$. 
The uncertainty of a given luminosity bin is estimated
assuming Poisson statistics from the relation:

\begin{equation}  
\delta\Phi(L)= \frac{\sqrt{N}}{\int_{L_{min}}^{L_{max}}  \int_{z_{min}(L)}^{z_{max}(L)}
\Omega(L,z)\,(\frac{dV}{dz})\,dz\, dL}.
\end{equation}

\noindent
 We also derive the luminosity function using the parametric Maximum
 Likelihood Method (ML; Tammann,Yahil   \& Sandage 1979).  We use a
 Schechter (1976) form for the luminosity function as this describes very well 
 the luminosity function in e.g. optical wavelengths  (Bingelli, Sandage \&
 Tammann 1988). Moreover, it has a strong theoretical footing 
 as it is derived from self-similar gravitational collapse models
 (Press \& Schechter 1974).  The Schechter function is expressed as: 
 
 \begin{equation}
 \Phi(L)=\phi_\star (L/L_\star)^{-\alpha} {\rm exp}(-L/L_\star) dL. 
 \end{equation}

\noindent 
In the above expression $L_\star$ denotes the characteristic
luminosity where the function above changes from a power-law with
slope $\alpha$ at the  faint-end to an exponential drop at brighter
luminosities. A likelihood function is constructed  as the product of
probabilities $P_i$ that a galaxy at redshift $z$ is detected with a
luminosity $L$.  Thus $P_i$ is  defined as the ratio of the number of
galaxies with luminosity between $L$ and $L+dL$ over the total number
observed, $P_i=\Phi(L)/\int^{\infty}_{L_{min(z)}} \Phi(L) dL $. Then
we maximise the sum  $\displaystyle \sum^{n} ln P_i$ by varying
$L_\star$ and $\alpha$.  The errors on $L_\star$ and $\alpha$
 are estimated from the $\delta L=0.5 $ regions around the 
 maximum likelihood fit. Since the normalization $\phi_\star$ of the
luminosity function cancels out in the  calculation above, we derive
$\phi_\star$ from  
 
 \begin{equation}
 \phi_\star= N_{gal}/ \int \int \Omega(L,z) \, \Phi(L)/\phi_\star \,
 dL \, dV/dz \, dz  
 \end{equation} 

\noindent
where $N_{gal}$ is the total number of galaxies in the survey and
$\Omega(L,z)$ is the solid angle of the X-ray survey available to a
source with luminosity $L$ at a redshift $z$, i.e. the area curve at
different flux limits. The uncertainty in $\phi_\star$ is approximated
by performing 200 bootstrap resamples of the data and then estimating
the 25th and 75th quartile around the median. For a Gaussian
distribution these correspond to the 68 per cent confidence level. 

We improve the statistical reliability of our luminosity function
estimates by combining our sample with 18 $z<0.22$  galaxies from the
CDF-N and CDF-S. As already discussed the CDF-N data are obtained from
Hornschemeier et al. (2003) by selecting a total of 10 sources with
0.5-8\,keV band detection. All of these systems have spectroscopic 
redshifts available. In the case of the CDF-S we select a total of 8
sources detected in the 0.5-2.0\,keV spectral band with $\log (f_x
/f_o)<-2$ from the catalogue presented by Giaconni et
al. (2002). Spectroscopic (total of 5) or photometric (total of 3)
redshifts are available from  Szokoly et al. (2004) and Zheng et
al. (2004) respectively.  

\begin{figure}
 \centerline{\psfig{figure=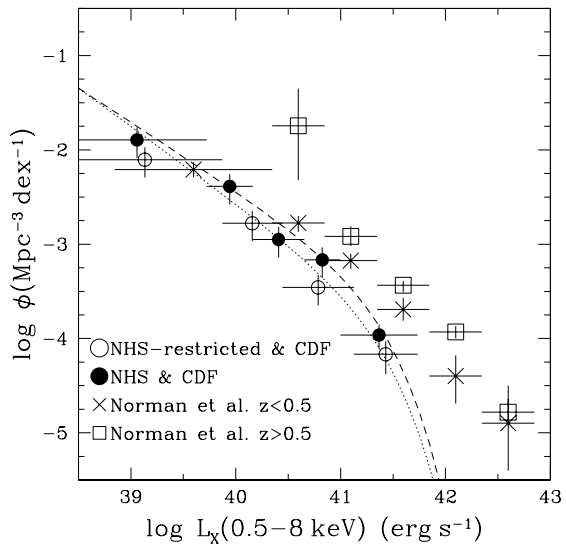,width=3in,angle=0}}
\caption
 { The local luminosity function derived using 
 the non-parametric method for the combined 
 NHS/CDF sample is denoted with filled (open) circles 
 in the case where we keep (exclude)
 the galaxies associated with the target.
 The dashed and dotted lines correspond to the 
 maximum likelihood determination in the above two cases 
 (NHS-restricted/CDF and NHS-total/CDF respectively).
 The luminosity function derived by Norman et al. (2004) 
 in two redshift bins is plotted for comparison.     
 }
\label{lf1}
\end{figure}

We estimate the luminosity function using the methods discussed
above for 3 different subsamples: (i) the NHS data alone, i.e. without
combining our sample with the CDF galaxies, (ii) both the NHS and the
CDF galaxies and (iii) combined NHS and CDF data after
excluding systems from our survey that are associated with the prime
target of a given \xmm pointing. The latter subsample is referred to
as the `restricted sample' and allows us to explore the sensitivity of
our results to the presence of group or cluster members within the
NHS. Figure \ref{lf1}
plots the local luminosity function for the subsamples (ii) and
(iii). For clarity we do not show the luminosity function for 
sample (i). We find that the luminosity function for the 
 restricted sample very closely resembles that of the total sample.
 Henceforth, we will be using the total sample (ii) in our analysis.  

\begin{figure}
 \centerline{\psfig{figure=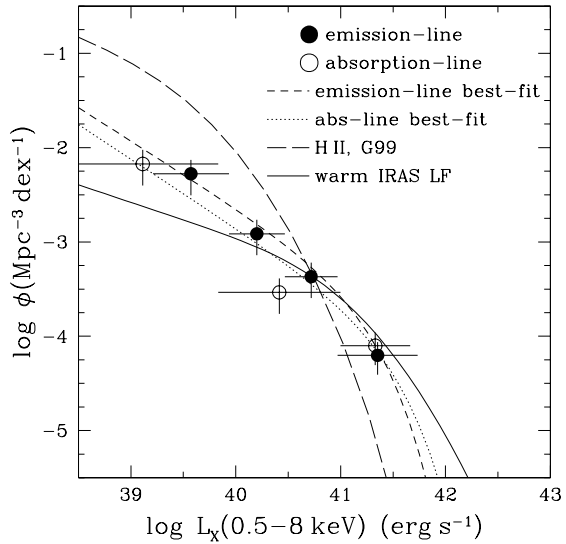,width=3in,angle=0}}
\caption
{ 
 The luminosity function derived using the 
 non-parametric method in the case of the 
 emission (filled circles) and absorption line (open circles) 
 galaxies. The short-dashed and dotted lines give the 
 corresponding  maximum likelihood Schechter form to these. 
 The long-dashed line represents the local $\phi(L_X)$ estimated
 indirectly from the optical star-forming 
 galaxy luminosity function (Georgantopoulos et al. 1999). The solid line 
 corresponds to the $\phi(L_X)$ derived from the 
 'warm' IRAS galaxy luminosity function of Takeuchi et al. (2003).   
 }
\label{lf2}
\end{figure}

The luminosity function derived above contains all galaxy types
i.e. both early and late.  Next, we attempt to explore the luminosity
function  for
different galaxy types using the combined NHS/CDF sample. Galaxies
with absorption optical lines are classified as early while systems
with narrow emission-lines or galaxies presenting both absorption and 
emission lines are grouped into the late type category. For systems
without optical spectra we use the best-fit SED estimated as a
by-product of the photometric redshift estimation for classification. 
There are 27 and 19 late and early type galaxies respectively. The
 results are shown in  Fig. \ref{lf2} and are  compared with the
 predicted star-forming X-ray galaxy luminosity function derived by
 Georgantopoulos, Basilakos \& Plionis (1999). This is estimated by
 convolving the optical star-forming luminosity function  with the
optical--to--X-ray luminosity relation. The optical luminosity
function has been derived  
 from the Ho et al. (1997) spectroscopic sample of galaxies
 whereas the optical--to--X-ray luminosity relation is taken from 
 the {\it Einstein} sample of Fabbiano et al. (1992).
 We also plot the X-ray luminosity function derived 
 by Norman et al. (2004) by convolving the  'warm' IRAS luminosity
 function (Takeuchi et al. 2003) with the  X-ray--to--far-infrared
 luminosity relation for star-forming galaxies 
(Ranalli et al. 2003).      
 
In Table \ref{lf} we summarise the best-fit parameters for the slope and the
break luminosity as well as the normalization derived from the maximum
likelihood method.  In the same table we give the X-ray
emissivity (luminosity per $\rm Mpc^3$)  
 
 \begin{equation}
 j_x=\int \Phi(L)~L~dL,
 \end{equation}

\noindent  
as well as the fractional contribution to the 
 0.5-8 keV X-ray background. The integrated galaxy X-ray flux is given by 
 
\begin{equation}
I=\frac{c}{4\pi H_{0}}\,
\int_{z1}^{z2}\frac{j_{x}~(1+z)^{p-\alpha_x}}{(1+z)(\Omega_m~(1+z)^3+\Omega_\Lambda)^{1/2}}\,
 \, dz.  
\end{equation}
 
\noindent
 We integrate all luminosities from $10^{38}\rm \, erg \, s^{-1}$  to
 infinity up to to a maximum redshift of $z=2$.  We 
 have assumed an energy spectral index of $\alpha_x=0.7$ (e.g. Zezas,
 Georgantopoulos \& Ward 1998).   The X-ray background intensity in
 the 0.5-8\,keV band is taken from Gendreau et al. (1995).     
 The X-ray flux sensitively depends on
 the assumed form of galaxy  evolution. Hopkins (2004) combined 
 the luminosity function information at many wavelengths, from radio 
 to X-rays and concluded that the luminosity density evolves 
 as  $(1 + z)^{p}$  with $p=3.3$
 for $z<1$, while for higher redshifts it appears to remain
 constant. Norman et al. (2004) find a luminosity evolution 
 consistent with p=2.7 at X-ray wavelengths up to their maximum
 redshift of $z\approx1$, close  to the value derived by Hopkins (2004).   
 In Table \ref{lf}, we give the contribution to the X-ray background
 ($I/I_{XRB}$) for both evolution indices. The errors for both $j_x$
 and $I/I_{XRB}$ are estimated in the same manner as the uncertainties
 in $\phi_\star$.

\begin{table*}
\footnotesize 
\begin{center} 
\begin{tabular}{l ccc ccc}
Sample    & $\log L_\star^1$ & $\alpha$ & $\phi_\star^2 $ & $j_x^3 $ & 
${I/I_{XRB}}^4$ & ${I/I_{XRB}}^5$ \\  
\hline 
NHS      &  $41.38_{-0.18}^{+0.23}$  &  $1.86^{+0.30}_{-0.29}$  &  
$3.40^{+1.61}_{-1.82}$ & $1.52_{-0.09}^{+0.16}$ & $0.17^{+0.05}_{-0.05}$ & 
 $0.12^{+0.02}_{-0.02}$  \\
NHS/CDF   &  $41.46_{-0.15}^{+0.18}$  &  $1.78^{+0.12}_{-0.12}$ &  
$2.54^{+3.13}_{-0.82}$ & $1.07_{-0.07}^{+0.06}$ & $0.15^{+0.01}_{-0.01}$ &
 $0.11^{+0.01}_{-0.01}$  \\
NHS/CDF restricted & $41.28_{-0.26}^{+0.21}$ & $1.80_{-0.18}^{+0.16}$ & 
$3.01_{-1.8}^{+2.3}$ & $0.89_{-0.04}^{+0.08}$ & $0.11^{+0.01}_{-0.01}$ &
 $0.08^{+0.01}_{-0.005}$ \\  
Emission  & $41.23^{+0.22}_{-0.17}$   & $1.71^{+0.17}_{-0.19}$ & 
$3.3^{+2.81}_{-2.19}$ & $0.65^{+0.04}_{-0.03}$ & $0.09^{+0.01}_{-0.01}$ &
 $0.06^{+0.004}_{-0.004}$ \\
Absorption& $41.68^{+0.33}_{-0.25}$  &  $1.81^{+0.16}_{-0.19}$  
&$0.58^{+0.88}_{-0.51}$ & $0.45^{+0.03}_{-0.03}$ & $0.06^{+0.004}_{-0.004}$ &
 $0.04^{+0.003}_{-0.003}$ \\
\hline 
\multicolumn{6}{l}{$^1$ in units $\rm erg~s^{-1}$; $^2$ in units $\rm  \times 
10^{-4} Mpc^{-3}~dex^{-1}$; $^3$ in units $\rm \times 10^{38} 
erg~s^{-1}~Mpc^{-3}$} \\
\multicolumn{6}{l}{$^4$ evolution index p=3.3; $^5$ p=2.7} \\
\end{tabular}
\end{center}
\caption{The luminosity function best-fit parameters}
\label{lf}
\end{table*}

\section{Discussion}
We use a total of 70 \xmm fields overlapping with the SDSS-DR2 to
compile a sample of 28 X-ray selected `normal' galaxies with
$z<0.22$. These systems have X-ray--to--optical flux ratios ($\log
(f_x / f_o) < -2$), luminosities ($\rm L_X < 10^{42} erg \,
s^{-1}$), X-ray colours and optical spectroscopic properties
(available for most of our sources) all suggesting X-ray emission
dominated by stellar processes (hot gas and X-ray binaries) rather
than accretion on a supermassive black hole.  Using this carefully
selected sample we construct the local ($z \la 0.2$)  X-ray
luminosity function of `normal' galaxies. Our \xmm survey nicely 
complements the deeper \chandra surveys in the coverage of the  $L_X -
z$ plane probing lower redshifts and higher luminosities. We combine
the two samples, exploiting the depth of \chandra and the wide areal
coverage of the NHS, to provide a `normal' galaxy sample totaling 46
systems at $z<0.22$. 

 We attempt to assess the efficiency of the  
 $\log(f_x/f_o)<-2$ criterion in selecting the most luminous normal galaxies. 
 We use the star-forming galaxy sample compiled by Zezas (2001) which
 comprises {\it ROSAT} PSPC observations of systems classified on the
 basis of high quality nuclear spectra from Ho el al. (1997).
 The above sample comprises 43 galaxies, detected by PSPC either as
 targets or serendipitously, spanning the luminosity range 
 $\rm L_X( 0.1 - 2.4 \, keV) \approx 4\times 10^{37}-
 3\times10^{41}$\,\lunits. We estimate the $\log(f_x/f_o)$ ratio from
 the 0.1-2.4 keV flux and the $B$-band magnitude. We find that no
 galaxy lies above the $\log(f_x/f_o)=-2$ cut, despite the fact  that
 highly luminous galaxies are included in the sample. Nevertheless, we
 note that the most X-ray luminous ($\approx 2\times10^{42}$\lunits)
 star-forming system known,  NGC3256 (Moran, Lehnert \& Helfand 1999),
 which is not included in  the Ho et al. (1997) sample,  has a
 relatively high X-ray--to--optical flux ratio, $\log (f_x/f_o) \approx
 -1.7$. This suggests that some very  X-ray luminous galaxies would
 evade our $\log (f_x/f_o) < -2$ criterion.
 This effect may be exacerbated at higher redshift. Indeed, in a scenario 
 where the $\log(f_x/f_o)$ increases with redshift (Hornschemeier 
 et al. (2003), the fraction of missed galaxies will be higher.  
 
 We further attempt to estimate the contamination of our sample by Low
 Luminosity AGN. We use the late-type galaxy sample of Shapley et
 al. (2001) comprising a total of 101 systems with $\log(f_x/f_o)
 <-2$. A number of these are classified AGNs, primarily using
 information from the optical spectra obtained by Ho et
 al. (1997). Note that we include only the  Seyfert and Liner1.9
 objects in the AGN class. We find 15 such objects  which satisfy the
 above criteria and this roughly translates to $\sim$15 per cent  
 contamination in the Shapley et al. sample. This may only represent 
 a lower limit as not all galaxies in Shapley et al. are common 
 with Ho et al. (1997) i.e. many systems do not 
 have good quality spectra. We note nevertheless, that even in the
 case where a small fraction of residual Low Luminosity  AGN is
 included in our sample  (because of the quality of the optical
 spectra),  this does not necessarily mean that the X-ray emission
 comes only from the AGN in these objects, e.g. Terashima \& Wilson
 (2003).  
 
The X-ray luminosity function of the combined sample with a median
redshift $z_{median}=0.076$ is compared in Figure \ref{lf1} with the
results at higher-$z$ of Norman et al. (2004). These authors derived
the first ever X-ray galaxy luminosity function, using data from the
combined CDF-North and South. Their sample probing redshifts up to
$z\approx1$ is split into two redshift bins with median  $z=0.26$ and
$z=0.66$  respectively. 
Inspection of  Fig. \ref{lf1} shows that their 'quasi-local'
$z<0.5$ luminosity function  is in good agreement with ours especially
at the faint end. At bright luminosities the  CDF luminosity function
is significantly higher than ours. This may suggest contamination of
the Norman et al. (2004) sample  by AGNs at bright luminosities. This
is not  highly unlikely, especially at luminosities brighter than
$10^{42}$\,\lunits,  since there is no optical spectroscopy available
for all the sources of Norman et al. (2004). Alternatively, we may be
witnessing evolution of the `normal' galaxy  luminosity
function.  The median redshift of the $z<0.5$  subsample of Norman et
al. is $z_{median}=0.26$ higher than our median redshift
$z=0.076$. For luminosity evolution of the form $(1+z)^{2.7}$  derived by
Norman et al. (2004), a source at $z=0.26$ is expected to become 1.5
times more luminous relative to $z=0.076$ . Moreover, we are excluding
from our analysis systems with X-ray to optical flux ratio $\log (f_x
/f_o)>-2$ and therefore, our  sample may be biased against X-ray
ultra-luminous star-forming galaxies,
 especially those  with $L_X \ga 10^{42} \rm \, erg
\, s^{-1}$ (see e.g. Moran et al. 1999).   
 Norman et al. (2004) use the log-norm functional form for fitting their 
 luminosity function. We note that such a form describes equally well our data;
 the fit yields $\Delta L\approx 1.4$ relative to the Schechter best-fit, 
 which can be considered however as only 
 a marginal improvement as the log-norm functional form 
 has an additional free parameter.
 In any case, the statistics are still limited and a detailed 
 comparison of the Schechter and log-norm functional forms 
 has to await till more data are accumulated.   

The luminosity function derived above encompasses both late and
early galaxy types. Figure \ref{lf2} presents the $\phi(L_X)$
estimates for these two classes separately. These are compared with
the local X-ray luminosity functions derived from (i) optically
selected star-forming galaxies (Georgantopoulos et al. 1999) and (ii)
warm IRAS galaxies (Norman  et al. 2004). The former largely
overestimates the number of emission-line systems at low luminosities
while the latter provides a better representation of the X-ray
luminosity function although it underpredicts the number of galaxies
with $L<L_\star$.      
 
 The optical luminosity function of early and late type-galaxies 
 has been derived by Madgwick et al. (2002) using a total of 75.000
 galaxies from  the 2dF Galaxy Redshift Survey  classified according  
 to their spectral properties. Four spectral types are defined ranging
 from passive absorption  line systems (earliest type) to actively
 star-forming galaxies (latest type).
 Madgwick et al.  (2002) find that the luminosity function of all four
 classes is well represented by a Schechter form with comparable
 $M^B_\star$ (within 0.4\,mag) and slopes that are getting
 steeper from early to late-type galaxies ($\alpha$ ranging from  0.54
 to 1.5). Interestingly, at X-ray wavelengths, the luminosity
 functions we derive for early and late type systems are comparable in
 both their shapes and normalizations, at odds with the results from 
 the optical regime. For example we note that  the slope of the X-ray
 luminosity function should be flatter than that at optical
 wavelengths for an X-ray--to--optical luminosity relation  steeper
 than linear (e.g. $L_X \propto L_B^{1.8}$; Fabbiano et
 al. 1992). Larger samples that the one used here are required to
 further explore this issue.   

Using the X-ray luminosity functions derived above we can provide the
most accurate yet estimates of the local galaxy X-ray emissivity 
(Table \ref{lf}). In the 0.5-8\,keV band this is estimated to be $\approx
10^{38}\, \rm erg \, s^{-1} \, Mpc^{-3}$ with about equal
contributions from early and late type systems. The emissivities
derived here are somewhat lower 
but still consistent within the uncertainties with those estimated by
Georgakakis et al. (2004a) via stacking analysis of 2dF galaxies.
Furthermore, adopting the star-formation rate evolution model of
Hopkins (2004) and Norman et al. (2004) we estimate the contribution
of galaxies to the X-ray background. This amounts to about 10-20 per
cent for all galaxies, up to a maximum redshift of $z=2$ with the
evolution truncated  at $z=1$. The exact fraction  depends on
the evolution index used and the sample used.  
 If we assume that the luminosity evolution continues up to z=2 we obtain
 contributions which are higher by about a factor of two.   
 We find a contribution to the X-ray background of  9 and 6
per cent  for emission and absorption line galaxies respectively using
the p=3.3 evolution model truncated at $z=1$. However, it is possible that 
the absorption line systems, associated with early-type galaxies, do not
present such strong evolution with cosmic  time (Lilly et
al. 1995). Assuming no evolution for these systems we assess that they
contribute about 2 per cent to the XRB. The fractions derived above
are higher than those in the CDF-North. For example Hornschemeier et
al. (2003) estimate  that 1-2 per cent of the 0.5-2\,keV XRB could
arise in normal galaxies. However, this is estimated by adding the
fluxes of optically selected galaxies in the CDF-North survey and
therefore should be considered as a lower limit as it does not take
into account the contribution of optically fainter systems.

\section{concluding remarks} 
The \chandra and \xmm missions opened a new window in the study of
distant galaxies  by providing  the first X-ray selected normal galaxy
sample. \xmm owing to its large field-of-view can constrain
efficiently the local ($z\la0.2$) X-ray galaxy  luminosity function.  
The \chandra deep fields probe normal galaxies with a median redshift
of $z\approx0.3$ (up to a maximum redshift of z=1) yielding information on the
evolution of the galaxies at X-ray wavelengths.  However, the peak of
the star-formation activity lies at even higher redshifts  which
remain beyond the reach of the current X-ray missions. These distant
galaxies reside at fluxes fainter than $10^{-17}$ \funits. This flux
regime can be accessed, and thus the study of galaxies at X-ray
wavelengths will only be furthered,  with the launch of high effective
area missions ($\rm >30 m^2$) combined with excellent positional
accuracy ($<2$ arcsec necessary to minimize confusion problems) 
such as the European Space Agency's mission {\it XEUS} .

\section{Acknowledgments}
 We are grateful to the anonymous referee for his/her suggestions
 which helped to improve substantially this paper.  
 This work is jointly funded by the European Union and the Greek
 Ministry of Development  in the framework of the programme 'Promotion of
 Excellence in Technological Development and Research', project
 'X-ray Astrophysics with ESA's mission XMM'. 
 We also acknowledge support from the Greek General Secretariat 
 for Research and Technology   programme 'Exploring 
 galaxies with NASA's {\it Chandra} X-ray mission'. We acknowledge the 
 use of data from the {\it XMM-Newton} Science Archive at VILSPA. 

 Funding for the creation and distribution of the SDSS Archive has
 been provided by the Alfred P. Sloan Foundation, the Participating
 Institutions, the National Aeronautics and Space Administration, the
 National Science Foundation, the U.S. Department of Energy, the
 Japanese Monbukagakusho, and the Max Planck Society. The SDSS Web
 site is http://www.sdss.org/. The SDSS is managed by the
 Astrophysical Research Consortium (ARC) for the Participating
 Institutions. The Participating Institutions are The University of
 Chicago, Fermilab, the Institute for Advanced Study, the Japan
 Participation Group, The Johns Hopkins University, Los Alamos
 National Laboratory, the Max-Planck-Institute for Astronomy (MPIA),
 the Max-Planck-Institute for Astrophysics (MPA), New Mexico State
 University, University of Pittsburgh, Princeton University, the
 United States Naval Observatory, and the University of Washington.

\end{document}